\documentclass{article}
\usepackage{authblk,amsmath,graphicx}

\usepackage{url}
\usepackage{array}

\usepackage{algorithm}
\usepackage{dsfont}

\usepackage{amssymb,color}

\title{A REGISTRATION ERROR ESTIMATION FRAMEWORK FOR CORRELATIVE IMAGING}

\author{Guillaume Potier $^{\star}$ \qquad Fr\'ed\'eric Lavancier $\dagger$ \qquad Stephan Kunne $^{\star}$ \qquad Perrine Paul-Gilloteaux $^{\star \ddagger}$ \thanks{Authors acknowledge the funding of the CROCOVAL project ANR-18-CE45-0015, PPG is part of the MicroPICell facility (BioGenouest), member of the national infrastructure France-Bioimaging (ANR-10-INBS-04) }}

\affil{$^{\star}$ Universit\'e de Nantes, CNRS, INSERM, l'institut du thorax, F-44000 Nantes, France.\\
$\dagger$ Universit\'e de Nantes, Laboratoire de Mathématiques Jean Leray, F-44000 Nantes, France.\\
$^{\ddagger}$ Universit\'e de Nantes, CHU Nantes, Inserm, CNRS, SFR Sant\'e, F-44000 Nantes, France}

\begin{document}
%
\maketitle
\begin{abstract}
Correlative imaging workflows are now widely used in bio-imaging and aims to image the same sample using at least two different and complementary imaging modalities. Part of the workflow relies on finding the transformation linking a source image to a target image. We are specifically interested in the estimation of registration error in point-based registration. We propose an application of multivariate linear regression to solve the registration problem allowing us to propose a framework for  the estimation of the associated error in the case of rigid and affine transformations and with anisotropic noise. These developments can be used as a decision-support tool for the biologist to analyze multimodal correlative images.
\end{abstract}
\section{Introduction}
\label{sec:intro}
\begin{figure}[htb]

\begin{minipage}[b]{0.5\linewidth}
  \centering
  \centerline{\includegraphics[width=5cm]{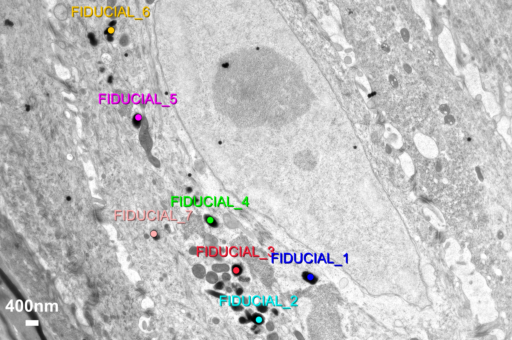}}
  \centerline{(a)}\medskip
\end{minipage}
\begin{minipage}[b]{0.5\linewidth}
  \centering
  \centerline{\includegraphics[width=5cm]{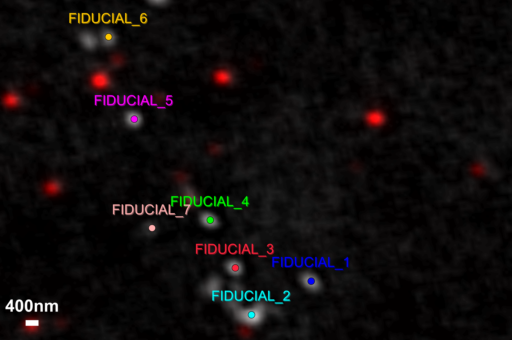}}
  \centerline{(b)}\medskip
\end{minipage}
\begin{minipage}[b]{0.5\linewidth}
  \centering
  \centerline{\includegraphics[width=5cm]{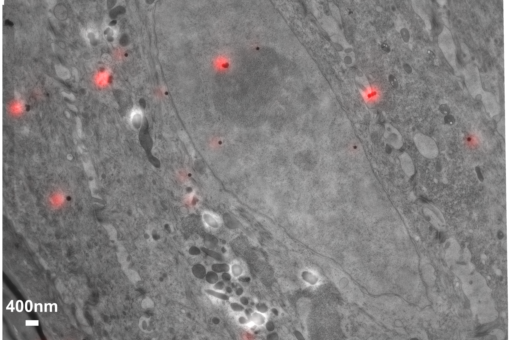}}
 \centerline{(c)}\medskip
\end{minipage}
\begin{minipage}[b]{0.5\linewidth}
  \centering
  \centerline{\includegraphics[width=5cm]{figures/leave_one_out_fractionperformedcorrelations2.png}}
  \centerline{(d)}\medskip
\end{minipage} \hfill
\begin{minipage}[b]{0.5\linewidth}
  \centering
  \centerline{\includegraphics[width=5cm]{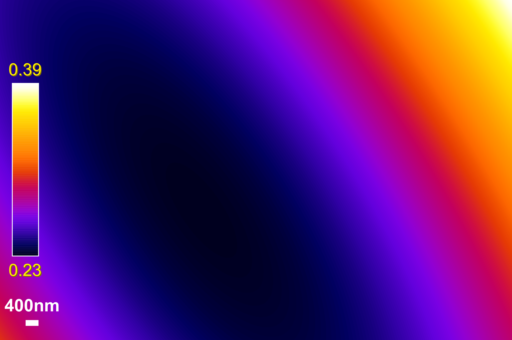}}
  \centerline{(e)}\medskip
\end{minipage} \hfill
\begin{minipage}[b]{0.5\linewidth}
  \centering
  \centerline{\includegraphics[width=5cm]{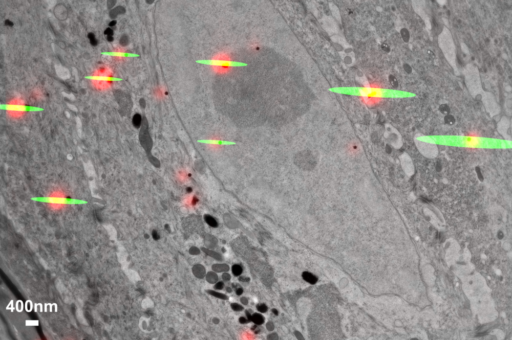}}
 \centerline{(f)}\medskip
\end{minipage}
\caption{Error representation in correlative imaging. (a) target (electron microscopy) with fiducials used; (b) registered source (fluorescence microscopy in red, bright field in inverted gray); (c) overlay of registered source and target, showing discrepancy of POIs (black and red); (d) Leave-one-out information as in \cite{Schorb2013}, only on fiducials; (e) Average expected error map as in \cite{paul-gilloteaux2017}; (f) Confidence ellipses at 95\% for POIs as we propose. Zoom for full resolution.}
\label{fig:error_information}
\end{figure}

Correlative microscopy has become ubiquitous in bio-medical research. 
The technique consists in observing the same sample under different complementary imaging modalities in order to gain more insight. 
The most known association is undoubtedly correlative light-electron microscopy (CLEM), but it is possible to combine other imaging modalities \cite{Walter2020}. 
The process includes a registration step in which the images of the different modalities are overlaid by estimating the transformation linking them. \\
In this work, we only consider point-based registration requiring a set of matched points, either fiducial points \cite{Kukulski2012}, anatomical or other natural landmarks \cite{Luckner2018} or cloud of points, which can be used in this context after a segmentation step by representing a contour or a surface by a cloud of points \cite{paul-gilloteaux2017}. 
All these usages can be modeled under a unique paradigm, that we will refer to as point-based registration in the following, and we will name points used for the registration as fiducial points, as opposed to points not used for the registration, called points of interest (POIs). 
The localization of fiducial points is prone to error, due to imaging resolution, method of localization or of sampling,... This will create a registration error for all points of the image \cite{Moghari2009,Fitzpatrick2001,Cohen2013}, having consequences on the overlaying of points of interest. 
There is a clear need in the bio-medical community to estimate errors in registration, as reflected by the reporting done in biological papers using correlative approaches (see Fig. \ref{fig:error_information}).
In correlative imaging, usual intensity metrics to assess the quality of the registration are usually not directly applicable due to the discrepancy of contents and scales. This is why generic cross-validation methods like leave-one-out are usually used to estimate the registration error as global values, based on fiducial points \cite{Schorb2013}. In \cite{paul-gilloteaux2017}, error map representation  based on the local average expected target registration error was proposed, using the statistical framework for rigid registration developed in  \cite{Fitzpatrick2001}. \\
The case of rigid transformations and estimated error distribution has been developed by Moghari and Abolmaesumi \cite{Moghari2009}. Unfortunately the rigid model can be too specific and fail to account for more general sample deformations. 
Cohen and Ober \cite{Cohen2013} developed an error-in-variable model solving for affine transformations. However due to their assumption of heteroscedastic noise it is necessary to estimate the noise distribution of each fiducial point which may not be possible from a practical point of view.  \\

In this work,  we provide a complete framework for the estimation of rigid and affine transformations providing at the same time error estimation for multidimensional image registration. We propose a new formulation for affine registration based on a multivariate linear regression modeling. In addition, we develop the maximum likelihood error estimation proposed in \cite{Moghari2009} under the rigid registration constraints.  These asymptotic expressions of the covariance matrix of the registration errors allow us to give a graphical representation of local accuracy as confidence ellipses at 95\%, as opposed to an estimation of the average error. Finally we demonstrate the application of these methods using simulations and real data.

\section{METHODS}

\subsection{General affine registration problem modeling}
\label{affine_registration_problem}
In our framework, in order to find the transformation linking a source image and a target image of respective dimension $r$ and $m$ (that are typically 2 or 3 for $2D$ or $3D$ images), each fiducial point of the source image is assigned a corresponding point on the target image.
We denote by $X$ the matrix containing the coordinates of the $n$ observed fiducial points $x_i\in\mathbb R^r$ ($i=1,\dots,n$) of the source image, stacked in rows so that $X$ is a $n\times r$ matrix. We similarly denote by $Y$ the matrix of size  $n\times m$ containing the $n$ observed fiducial points $y_i\in\mathbb R^m$ of the target image.

Let $Z=[\mathds{1}\ X]$ be the matrix of size $n\times (r+1)$ where all components of the first column are 1. 
If the transformation between $X$ and $Y$ was affine, we would have exactly $Y=Z\beta$ for some matrix $\beta$ of size $(r+1)\times m$. In practice of course this relation is just an approximation of the true transformation and a modeling error has to be added. Another error, the so-called localization error,  comes from the resolution limit of the images and the possible presence of noise, that make the coordinates of the fiducial points inaccurate and so the theoretical relation -- even it was true -- imprecise for the observed fiducial points. 
However, $X$ and $Y$ gather the observed locations of fiducial points (and not the theoretical locations of them) and thus already include the localization errors. For this reason, we do not deal with an errors-in-variables model as studied in \cite{Cohen2013}. 
We consider the following model relating $Y$ and $X$: 
\begin{equation}\label{model1}
\underset{(n \times m)}{Y} = \underset{(n \times (r + 1))}{\begin{bmatrix} \mathds{1} & X \end{bmatrix}} \times \underset{((r + 1) \times m)}{\beta} + \underset{(n \times m)}{\epsilon},
\end{equation}
where $\epsilon$ is a random matrix error of size $n \times m$. Accordingly, each row $\epsilon_i$ of $\epsilon$ is a vector of dimension $m$ corresponding to the random modeling error associated to the $i$-th pair of observed fiducial points $(x_i,y_i)$.

We assume that the $\epsilon_i$'s are independent, centered and follow the same multidimensional Gaussian distribution in $\mathbb R^m$ with unknown variance $\Sigma$ of size $m\times m$, i.e. $\epsilon_i\sim  \mathcal{N}_{m}(0, \Sigma)$ for all $i=1,\dots,n$. We do not assume any particular constraints on $\beta$ and $\Sigma$ apart, for the latter, to be a well-defined covariance matrix. This means that our setting includes as particular cases: i) rigid transformation, the case where the affine transformation reduces to a translation combined with a rotation, as considered for instance  in \cite{Fitzpatrick2001,Moghari2009,paul-gilloteaux2017}; ii) isotropic errors, when $\Sigma$ is proportional to the identity matrix; and iii) uncorrelated errors in each direction, the case when $\Sigma$ is diagonal. Our general setting thus accounts for a general affine transformation and a general error that can be anisotropic and correlated along the different directions of the image. In this setting, $\beta$ and $\Sigma$ are unknown.
Our goal is then to be able to predict the location $y_0$ of a POI in the target image associated to $x_0$ in the source image, and to get a confidence region for the registration error, here the error associated to this prediction. We provide in the following sections analytic expressions of these confidence regions, making their computation fast and reliable. We consider the general affine model first and then discuss the more constrained rigid model. 

\subsection{Error estimation for an affine model}
\label{model_affine}
The affine model \eqref{model1} is a multivariate linear regression model, see for instance \cite[Chapter 7]{johnson2007}. The maximum likelihood  estimators of $\beta$ and $\Sigma$ are respectively  $\hat{\beta} = (Z^{'}Z)^{-1}Y$ and $\hat{\Sigma} =\hat{\epsilon}'\hat{\epsilon}/n$, where $\hat{\epsilon} = Y - Z\hat{\beta}$ is the residuals matrix and where we recall that $Z$ denotes the matrix $[\mathds{1}\ X]$. 

Once the model is fitted, we can predict the position $y_0\in\mathbb R^m$ in the target image of a point of interest $x_0\in\mathbb R^r$ observed in the source image. Let $z_{0} = (1, x_{0}')'\in\mathbb R^{r+1}$, then according to the affine model \eqref{model1}, $y_0=\beta' z_0+\epsilon_0$ for an error $\epsilon_0\sim \mathcal{N}_{m}(0, \Sigma)$ independent of all $\epsilon_i$'s, $i=1,\dots,n$, meaning that $\epsilon_0$ is independent of $\epsilon$. The prediction of $y_0$ is then $\hat y_0= \hat{\beta}^{'}z_{0}$ and the prediction error is:
$$y_{0} - \hat y_0 = (\beta - \hat{\beta})^{'}z_{0} + \epsilon_{0}.$$
This error is distributed as a centered Gaussian distribution in $\mathbb R^m$ with variance $(1 + z_{0}^{'}(Z^{'}Z)^{-1}z_{0}) \Sigma$. Plugging the estimate $\hat\Sigma$, we obtain that the  confidence ellipsoidal region $\mathcal E(y_0)$ of $y_0$ at the significance level $\alpha$ is given by the inequality, see \cite{johnson2007}: for all $y\in\mathcal E(y_0)$, 
\begin{multline*}
(y - \hat{\beta}^{'}z_{0})^{'}\left(\frac{n}{n - r - 1}\hat{\Sigma}\right)^{-1}(y - \hat{\beta}^{'}z_{0}) 
\\ \leq (1 + z_{0}^{'}(Z^{'}Z)^{-1}z_{0})\left(\frac{m(n - r - 1)}{n - r -m}\right)F_{m, n - r - m}(1 - \alpha),
\end{multline*}
where $F_{m, n - r -m}(1 - \alpha)$ denotes the $(1 - \alpha)^{th}$ percentile of a Fisher's law with parameters $m$ and $n - r - m$. By construction, $\mathcal E(y_0)$ has a probability $1-\alpha$ to contain the unknown location $y_0$. This confidence region is available analytically, making its computation fast and accurate, and importantly, it depends on the source location $x_0$ through $z_0$.

\subsection{Error Estimation under the constraints of the rigid model}
\label{model_rigid}

Rigid models are considered in \cite{Fitzpatrick2001} and \cite{Moghari2009}, among others. They correspond to the  particular case of \eqref{model1} where $r=m$ and $\beta'=[t\  R_\theta]$ is composed of a translation vector $t\in\mathbb R^m$ and of a rotation matrix $R_\theta$ of size $m\times m$. Here $\theta$ is the parameter of size $m(m-1)/2$ characterizing the rotation. The transformation matrix $\beta$ of a rigid motion thus involves $m(m+1)/2$ free parameters against $m(m+1)$ in the affine case. 
As in the affine case we assume that each error $\epsilon_i$ independently follows a $\mathcal N_m(0,\Sigma)$ and a maximum likelihood procedure can be used to estimate the unknown parameters $\beta$ (under the above constraints) and $\Sigma$. In the general case, this optimization problem amounts to find $t$, $\theta$ and $\Sigma$ that maximize 
\begin{multline*}
n\log\left(\frac{1}{\sqrt{(2\pi)^{m}|\Sigma|}}\right) \\ - \frac{1}{2}\sum_{i = 1}^{n} [y_{i} - (R_{\theta} x_{i} + t)]'\Sigma^{-1}[y_{i} - (R_{\theta} x_{i} + t)].
\end{multline*}
This problem, discussed in \cite{Moghari2009}, does not admit a closed-form solution and some numerical optimization procedure are needed to solve it in the general case. However, in the particular case where the errors are assumed to be isotropic, meaning that $\Sigma$ is proportional to the identity matrix,  this problem reduces to a constrained least squares optimization problem known as the orthogonal Procrustes problem. An analytic solution in this setting is known, see    \cite{Schonemann1966,Kabsch1978}. In our implementation, we use the latter solution in the isotropic case, while we use a numerical solver in the general case.  

For a rigid model, the prediction of the target point $y_0=R_{\theta} x_0 + t + \epsilon_0$ associated to a new source point $x_0$ is $\hat y_0= R_{\hat\theta} x_0 + \hat t$, where $\hat\theta$ and $\hat t$ are the maximum likelihood estimates. The registration error is thus $y_0-\hat y_0= e(\hat t, \hat \theta) + \epsilon_0$ where $e(\hat t, \hat \theta)=(R_\theta - R_{\hat\theta})x_0 + (t-\hat t)$ is an estimation error independent of $\epsilon_0$ (as  $\epsilon_0$ is independent of $\epsilon$). Accordingly, the covariance matrix of the registration error is  $\Sigma_e + \Sigma$ where $\Sigma_e$ is the covariance matrix of $e(\hat t, \hat \theta)$.  By the law of propagation of uncertainties, $\Sigma_{e}$ is asymptotically equal to 
$J_{e} \Sigma_{\hat t, \hat \theta} J_{e}^{'}$ where $J_e$ is the Jacobian matrix of the function $e$ and $\Sigma_{\hat t, \hat \theta}$ is the covariance matrix of $(\hat t, \hat \theta)$. 
Denoting by $q=m(m-1)/2$ the dimension of $\theta$, $J_e$ is the $m\times (m+q)$ matrix  given by 
$$J_e=-\left(I_m \ |\ \partial_1 R_{\hat\theta} x_0 \ |\ \dots \ |\ \partial_q R_{\hat\theta} x_0\right),$$ where $I_m$ is the identity matrix of size $m$ and $\partial_k R_\theta$ denotes the (element-wise) derivative of the matrix $R_\theta$ with respect to the element $\theta_k$ ($k=1,\dots,q$) of $\theta$.
As to $\Sigma_{\hat t, \hat \theta}$, it is asymptotically equivalent to the inverse Fisher information matrix of the model, in agreement with the asymptotic efficiency of the maximum likelihood estimator $(\hat t, \hat \theta)$ \cite{Rao1992}, i.e.
$\Sigma_{\hat t, \hat \theta}\sim \mathcal I^{-1} (t,\theta,\Sigma; X)$ where $\mathcal I$ is the block matrix 
\begin{equation*}\label{fisher}
\mathcal I (t,\theta,\Sigma; X) =\left(\begin{array}{c@{}|@{} c}
\mathcal I_{tt} & $\quad $  \begin{matrix} \mathcal I_{t\theta_1} & \cdots & \mathcal I_{t\theta_q} \end{matrix} \\ \hline
\begin{matrix} \mathcal I_{t\theta_1}' \\ \vdots \\ \mathcal I_{t\theta_q}'\end{matrix} &  \mathcal I_{\theta\theta}
\end{array}\right).
\end{equation*}
The computation of $\mathcal I$ is approximated by linearisation in \cite{Moghari2009} when $m=3$ and under the hypothesis that $\theta$ is small, but exact formulas can be derived. Specifically, we have 
\begin{align*}
\mathcal I_{tt} &= n\Sigma^{-1},\\
\mathcal I_{t\theta_k} &= \sum_{i=1}^n \Sigma^{-1}\partial_k R_\theta x_i, \quad k=1,\dots,q,\\
\mathcal I_{\theta\theta} &=\left[  \sum_{i=1}^n x_i'\partial_k R_\theta\Sigma^{-1}\partial_l R_\theta x_i\right]_{k,l =1,\dots,q}.
\end{align*}
In our implementation, we use these exact formulas. In the end, the asymptotic approximation of the variance of the registration error in the rigid case writes $J_{e} \hat\Sigma_{\hat t, \hat \theta} J_{e}^{'} + \hat \Sigma$, where $\hat\Sigma_{\hat t, \hat \theta}=\mathcal I (\hat t,\hat \theta,\hat\Sigma; X)$. Based on this variance, asymptotic confidence ellipsoidal region can be constructed, exactly as carried out in the affine case.

\section{RESULTS}
\label{results}
\subsection{Accuracy of confidence region and of registration}
In addition to test on real data as demonstrated on Figure \ref{fig:error_information}, simulations were performed with rigid and affine transformations and 10, 25 and 100 fiducial points. Here we present the results for $95^{th}$ percentile of the used Fisher's law, but any $\alpha$ could be used. When we compute a 95\% prediction ellipse then theoretically the estimated point lies within the ellipse 95\% of the time. Simulations are repeated 10 000 times. Prediction ellipses are computed for 100 points of interest drawn from a uniform distribution in the square of size 1024x1024 pixels. Fiducial points were drawn from a Gaussian distribution centered at the point (256, 256) with isotropic variance equal to 500 in each direction.\\
The coverage rate is defined as the number of times the noisy point of interest lies within the computed prediction ellipse, divided by the number of iterations. This statistic is used to check whether the computed confidence ellipses are correct. The mean area of the prediction ellipses gives an indication of the size of the estimated error and allows to assess the quality of the registration.
\begin{figure}[htb]
\begin{minipage}[b]{1.0\linewidth}
  \centering
  \centerline{\includegraphics[width=12cm]{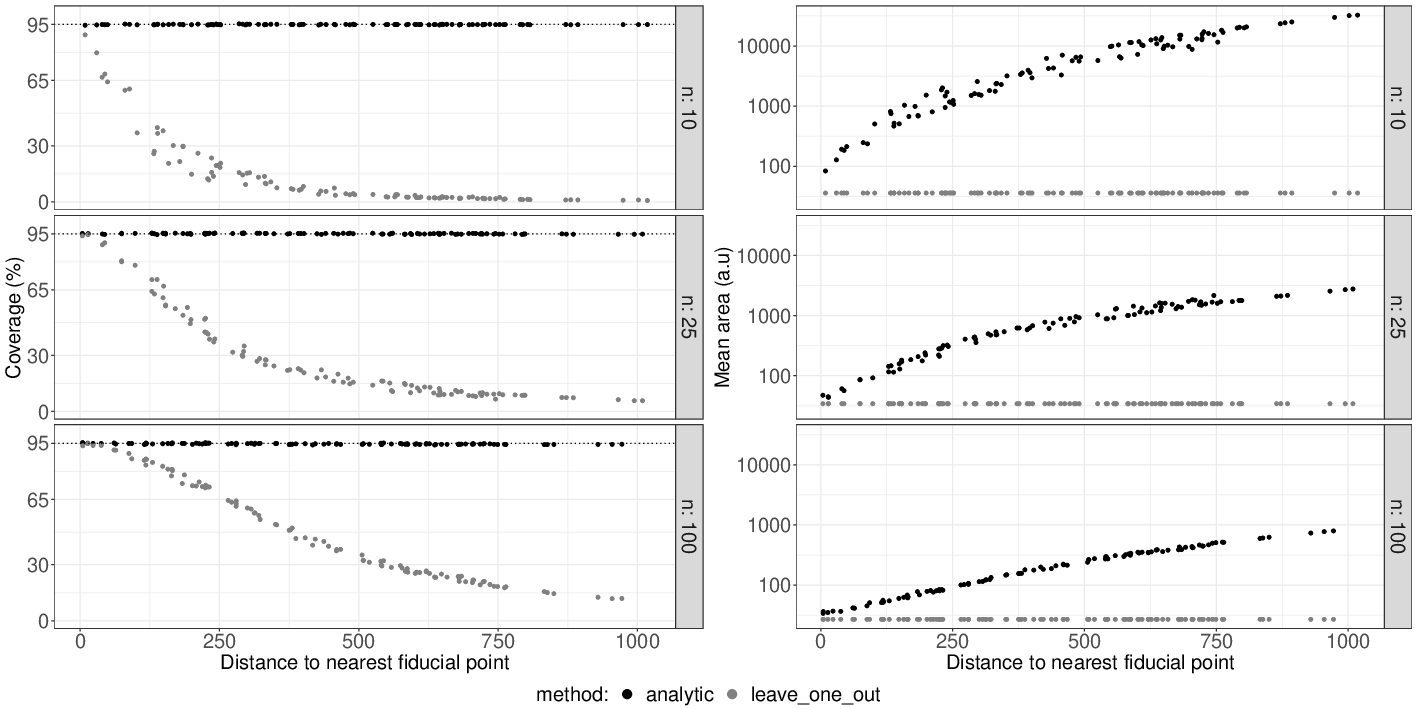}}
\end{minipage}
\caption{Coverage and mean area of prediction ellipses at different test points for affine model computed with analytic and leave-one-out methods using 10, 25, or 100 fiducial points under an affine transformation}
\label{fig:loo}
\end{figure}
Table \ref{tbl:rigid} compares the distribution of the coverage rate of 95\% prediction ellipses computed with the analytic methods of the rigid and affine models under a rigid or affine transformation. Coverage rates distribution lies at 95\% with high precision and accuracy. Distribution of coverage rates for the rigid model is shifted above 95\% because the model assumes an asymptotic minimum variance. The distribution converges towards 95\% when the number of fiducial points is increased. The affine model can still produce good estimations with rigid transformation and is more robust than the rigid model.
We use the same variant of leave-one-out method as described in \cite{Schorb2013}. In this flavor, the prediction area is the disk whose radius is the 95th quantile of the measured errors. Results presented in figure \ref{fig:loo} show that with our method  the coverage rate is correctly constant at 95\% and the mean area of predicted ellipses increase with the distance to the nearest fiducial point. The leave-one-out method underestimates the registration error since the registration error estimated by the leave-one-out method is the same at any location. As shown in figure \ref{fig:error_information}(f), our confidence ellipses can be a useful indication for the biologist in order to associate unknown structures (here q-dots used for demonstration as POIs, when other structures were used as fiducials), and complementary of the error maps proposed Fig.\ref{fig:error_information}(e). \ref{fig:error_information}.
\begin{table}[h]
  \caption{\label{tbl:rigid}Distribution of the coverage rate of 95\% prediction ellipses (Cov\%, target value is 95\%) for rigid and affine models (model) computed with analytic method under rigid and affine transformations (Transfo) using 10 and 100 fiducial points}
  \begin{center}
\begin{tabular}{ |c|c|c|c| }
\hline
Transfo & Model & Cov\% n=10 & Cov\% n=100 \\
 &  & mean/STD & mean/STD \\
\hline
rigid & rigid & 99.35 / 0.08&95.68 / 0.21\\
rigid & affine & 94.98 / 0.14 &94.57 / 0.22\\
affine & rigid & 3.53 / 17.51&2.00 / 14.07\\
affine & affine & 95.05 / 0.21&94.62 / 0.34\\
\hline
\end{tabular}
\end{center}
\end{table}
\subsection{Testing the influence of attenuation bias in our bio-medical context}
\begin{table}[h]
  \caption{\label{tbl:attenuation_bias}Distribution of coverage rates (min, mean, STD, max, 99\% confidence interval for mean coverage rate) for 95\% prediction ellipses obtained by simulation (model affine/transformation affine) including attenuation bias under the condition of microscopy for 10, 25 and 100 fiducial points}
  \begin{center}
\begin{tabular}{ |c|c|c|c|c|c| }
\hline
n & min & mean & STD & max & CI 99\% of mean \\
\hline
10 & 94.54 & 94.94 & 0.13 &  95.31 & $\begin{bmatrix}94.90,94.97\end{bmatrix}$\\
25 & 94.61 & 95.07 & 0.21 &  95.48 & $\begin{bmatrix}95.02,95.13\end{bmatrix}$\\
100 & 94.08 & 94.68 & 0.28 &  95.33 & $\begin{bmatrix}94.61,94.75\end{bmatrix}$\\
\hline
\end{tabular}
\end{center}
\end{table}
As explained in section \ref{affine_registration_problem}, our observations $X$ and $Y$ in model (\ref{model1}) include the localization error, and assume that both fiducials and POIs observations contains noise. If we change the paradigm such that $X$ does not include a localization error (which could indeed be negligible in case of very different scale), we are back to an errors-in-variable problem and we could suffer from the attenuation bias. 
In a microscopy context, localization are often sought to have a subpixelic resolution. Taking into account a broader range of case, assuming a 3 pixels radius of error in the localization sounds a reasonable assertion for fiducial points in the image with less resolution. 
So if we model the localization error in pixels according to $\mathcal{N}\left(0, I_2 \right)$ then the localization error lies within +/- 3 pixels. Results presented in table \ref{tbl:attenuation_bias} clearly indicate there is a bias when we assume both noise since computed 99\% confidence intervals for the mean coverage rate does not includes 95\% which is the theoretical value. However the bias never exceeded 1\% deviation from the theoretical value. We conclude the attenuation bias in our context is negligible. 

\section{DISCUSSION AND CONCLUSION}
In this article we described the problem of point-based registration as a linear least squares regression problem and propose tools for registration error estimation. We show through simulations that the registration by linear regression in the affine case is more robust than the rigid method. We demonstrate that cross-validation registration error estimation like leave-one-out may be unreliable because it underestimates the registration error. We provide an implementation of registration and error estimation under rigid and affine models in 2D and 3D as an ICY plugin \cite{deChaumont2012}. This method provides analytic registration error estimation through prediction ellipses which is a visual and intuitive way of assessing the physical matching of two unknown structures and the registration quality. Furthermore our source code is released and available at \url{https://github.com/anrcrocoval/ec-clem} with video example on real data  \url{https://www.youtube.com/watch?v=Rz1_MLqn6-k}. Other registration methods can be used to account for non-linear deformations in biological samples \cite{Holden2008}. Due to the non-linear property, registration error is difficult to estimate, but our method could be directly extended to a local affine registration framework to take into account local deformation.

\bibliographystyle{plain}
\bibliography{references}

\begin{thebibliography}{10}

\bibitem{Cohen2013}
E.~A.~K. Cohen and R.~J. Ober.
\newblock Analysis of point based image registration errors with applications
  in single molecule microscopy.
\newblock {\em IEEE Trans Signal Processing}, (61(24)):6291–6306, 2013.

\bibitem{deChaumont2012}
Fabrice de~Chaumont, Stéphane Dallongeville, Nicolas Chenouard, Nicolas
  Hervé, Sorin Pop, Thomas Provoost, Vannary Meas-Yedid, Praveen Pankajakshan,
  Timothée Lecomte, Yoann Le~Montagner, Thibault Lagache, Alexandre Dufour,
  and Jean-Christophe Olivo-Marin.
\newblock Icy: an open bioimage informatics platform for extended reproducible
  research.
\newblock {\em Nature Methods}, 9(7):690--696, July 2012.

\bibitem{Fitzpatrick2001}
J.~Fitzpatrick and Jay West.
\newblock The distribution of target registration error in rigid-body,
  point-based registration.
\newblock {\em IEEE transactions on medical imaging}, 20:917--27, 10 2001.

\bibitem{Holden2008}
M.~{Holden}.
\newblock A review of geometric transformations for nonrigid body registration.
\newblock {\em IEEE Transactions on Medical Imaging}, 27(1):111--128, Jan 2008.

\bibitem{johnson2007}
Richard~Arnold Johnson and Dean~W Wichern.
\newblock {\em Applied multivariate statistical analysis}.
\newblock Upper Saddle River, NJ: Prentice hall, 6th edition, 2007.

\bibitem{Kabsch1978}
W.~Kabsch.
\newblock {A discussion of the solution for the best rotation to relate two
  sets of vectors}.
\newblock {\em Acta Crystallographica Section A}, 34(5):827--828, Sep 1978.

\bibitem{Kukulski2012}
W.~Kukulski, M.~Schorb, S.~Welsch, A.~Picco, M.~Kaksonen, and J.A.G. Briggs.
\newblock Precise, correlated fluorescence microscopy and electron tomography
  of lowicryl sections using fluorescent fiducial markers.
\newblock {\em Methods in Cell Biology}, 111:235--257, 2012.
\newblock cited By 74.

\bibitem{Luckner2018}
Manja Luckner, Steffen Burgold, Severin Filser, Maximilian Scheungrab, Yilmaz
  Niyaz, Eric Hummel, Gerhard Wanner, and Jochen Herms.
\newblock Label-free 3d-clem using endogenous tissue landmarks.
\newblock {\em iScience}, 6:92 -- 101, 2018.

\bibitem{Moghari2009}
M.~H. {Moghari} and P.~{Abolmaesumi}.
\newblock Distribution of target registration error for anisotropic and
  inhomogeneous fiducial localization error.
\newblock {\em IEEE Transactions on Medical Imaging}, 28(6):799--813, June
  2009.

\bibitem{paul-gilloteaux2017}
Perrine Paul-Gilloteaux, Xavier Heiligenstein, Martin Belle, Marie-Charlotte
  Domart, Banafshe Larijani, Lucy Collinson, Graça Raposo, and Jean Salamero.
\newblock {eC}-{CLEM}: flexible multidimensional registration software for
  correlative microscopies.
\newblock {\em Nature Methods}, 14(2):102--103, February 2017.

\bibitem{Rao1992}
C.~Radhakrishna Rao.
\newblock {\em Information and the Accuracy Attainable in the Estimation of
  Statistical Parameters}, pages 235--247.
\newblock Springer New York, New York, NY, 1992.

\bibitem{Schonemann1966}
Peter~H. Sch{\"o}nemann.
\newblock A generalized solution of the orthogonal procrustes problem.
\newblock {\em Psychometrika}, 31(1):1--10, 1966.

\bibitem{Schorb2013}
M.~Schorb and J.~A. Briggs.
\newblock Correlated cryo-fluorescence and cryo-electron microscopy with high
  spatial precision and improved sensitivity.
\newblock {\em Ultramicroscopy}, (143):24--32, 2014.

\bibitem{Walter2020}
Andreas Walter, Perrine Paul-Gilloteaux, Birgit Plochberger, Ludek Sefc, Paul
  Verkade, Julia~G. Mannheim, Paul Slezak, Angelika Unterhuber, Martina
  Marchetti-Deschmann, Manfred Ogris, Katja Bühler, Dror Fixler, Stefan~H.
  Geyer, Wolfgang~J. Weninger, Martin Glösmann, Stephan Handschuh, and Thomas
  Wanek.
\newblock Correlated multimodal imaging in life sciences: Expanding the
  biomedical horizon.
\newblock {\em Frontiers in Physics}, 8:47, 2020.

\end{thebibliography}

\end{document}